\documentclass[twocolumn,aps,prl,showpacs,floatfix,superscriptaddress]{revtex4}

\usepackage{amsmath}    
\usepackage{graphicx}   
\usepackage{epsfig}
\usepackage{verbatim}   
\usepackage{color}      
\usepackage{subfigure}  
\usepackage{hyperref}   

\begin{document}

\def\eps{\varepsilon}
\def\epsCr{\varepsilon_\mathrm{cr}}

\title{Breakup of small aggregates driven by turbulent hydrodynamic stress}
\author{Matthaus U. Babler} 

\affiliation{{Dept. of Chemical Engineering and Technology, Royal Institute of Technology, 10044 Stockholm, Sweden}}

\author{Luca Biferale}
\affiliation{Dept. of Physics and INFN, University of Rome Tor Vergata, Via della Ricerca Scientifica 1, 00133 Roma, Italy}

\author{Alessandra S. Lanotte}
\affiliation{ISAC-CNR, Str. Prov. Lecce-Monteroni, and INFN, Sez. Lecce, 73100 Lecce, Italy}

\begin{abstract}
Breakup of small solid aggregates in homogeneous and isotropic
turbulence is studied theoretically and by using Direct Numerical
Simulations at high Reynolds number, $Re_{\lambda} \simeq 400$. We
show that turbulent fluctuations of the hydrodynamic stress along the
aggregate trajectory play a key role in determining the aggregate mass
distribution function. Differences between turbulent and laminar flows
are discussed. A novel definition of the fragmentation rate is
proposed in terms of the typical frequency at which the hydrodynamic
stress becomes sufficiently high to cause breakup along each
Lagrangian path. We also define an Eulerian proxy of the real
fragmentation rate, based on the joint statistics of the stress and
its time derivative, which should be easier to measure in any
experimental set-up. Both our Eulerian and Lagrangian formulations
define a clear procedure for the computation of the mass distribution
function due to fragmentation. Contrary, previous estimates based only
on single point statistics of the hydrodynamic stress exhibit some
deficiencies. These are discussed by investigating the evolution of an
ensemble of aggregates undergoing breakup and aggregation.
\end{abstract}
\date{}  \maketitle 

\begin{figure}[t]
\includegraphics[width=75mm]{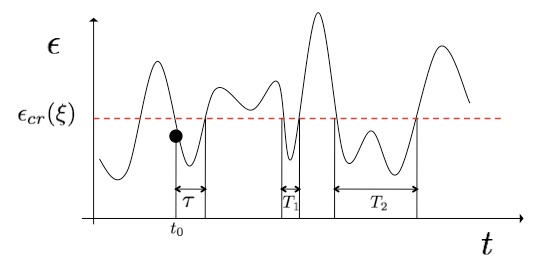}
\vspace{-0.4cm}
\caption{\label{fig:1} Pictorial evolution of the energy dissipation
  $\eps$ along an aggregate trajectory. Starting to record the stress
  at time $t_0$, the exit-time $\tau$ and the diving-times $T_1,\ T_2,
  \ldots$ for a given threshold $\epsCr$ are shown.}
\end{figure}

\begin{figure}
\includegraphics[width=75mm]{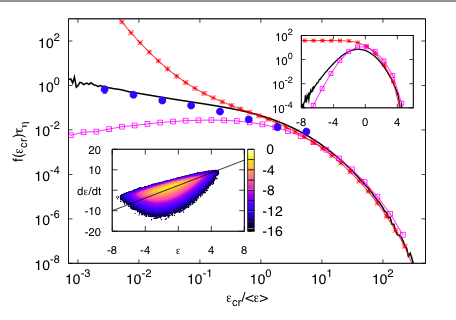}
\caption{The normalized fragmentation rate $f_{\epsCr} \tau_{\eta}$
  versus the normalized energy dissipation
  $\epsCr/\langle\eps\rangle$. Filled circles are the definition
  (\ref{eq:exit}), measured up to thresholds where statistical
  convergence of exit-times is obtained; the solid line is
  $f_{\epsCr}^E$. Squares are $f_{\epsCr}^{I}$, while crosses are
  $f_{\epsCr}^{II}$. (Bottom inset) Joint distribution $p_2(\eps,\dot
  \eps)$. The continuous line is the dimensional estimate $\dot \eps
  \sim \eps/\tau_{\eta}(\eps)$. (Top inset) Numerator of
  Eqs.~(\ref{eq:loginov}) -- (\ref{babler}); curves colors are the
  same of the main figure. \label{fig:2}}
\end{figure}

\begin{figure}
\includegraphics[width=82mm]{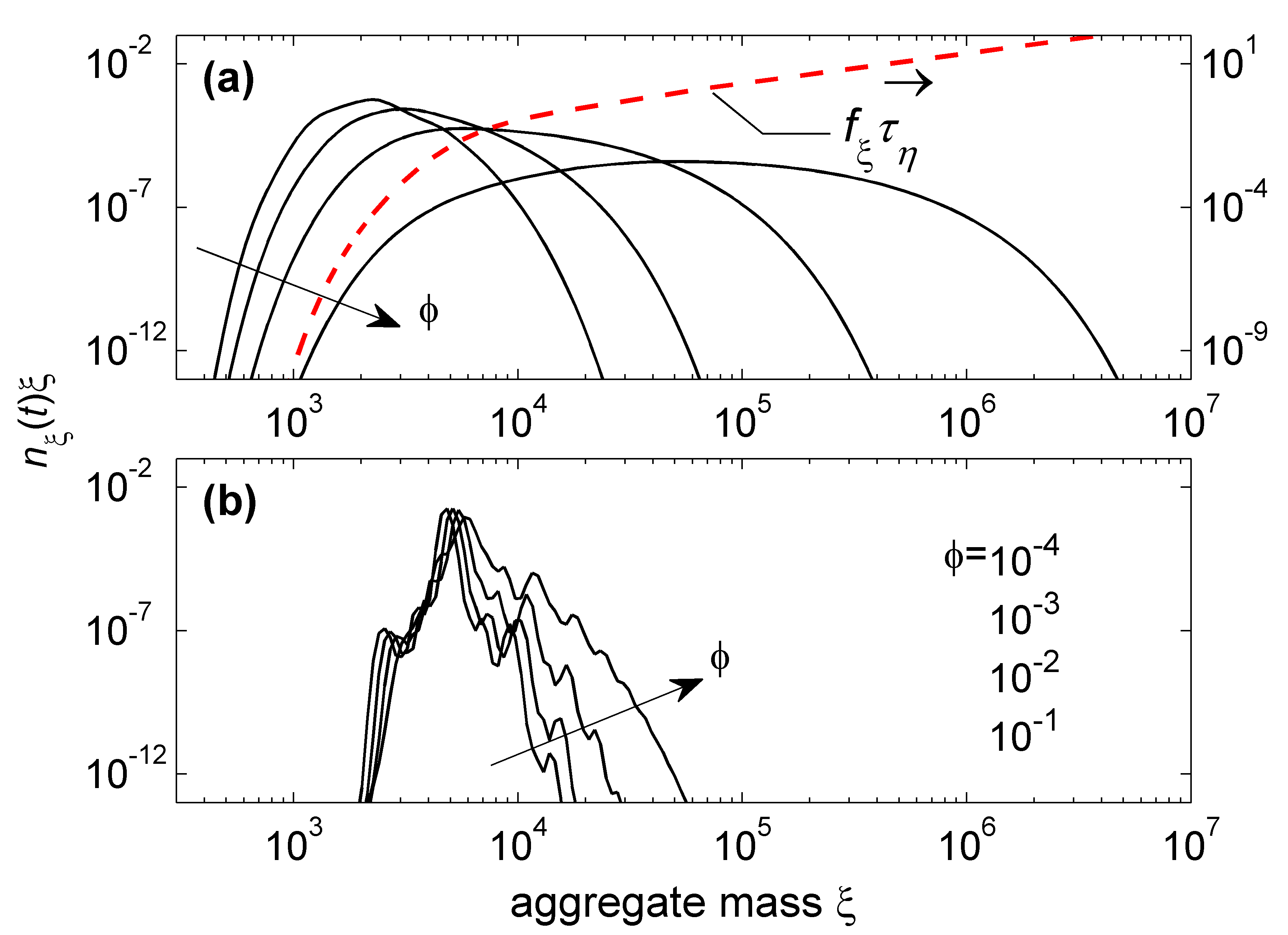}
\caption{\label{fig:4} Stationary mass distribution $n_\xi(t)$ with
  $d_f=2.4$ and $q=0.36$, for different solid volume fractions
  $\phi$. Distributions obtained in the turbulent flow at
  $Re_{\lambda}\simeq400$ (a) are compared to those of the laminar
  cases (b), with the uniform shear rate $\eps_{cr}^{lam}=\langle
  \epsilon \rangle$. In (a), we plot $f_\xi$ (dashed line) as given by
  Eq.~(\ref{eq:loginov}), assuming a power-law behavior in the limit
  of large aggregate mass.}
\end{figure}

Turbulence has a distinct influence on the aggregation of colloidal
and aerosol particles. It not only enhances the rate of collision
among particles, i.e. by inducing high velocity differences and
preferential concentration within the particle field
\cite{BiferaleJFM2010a,BiferaleJFM2010b}, but also it creates
hydrodynamic stress that can cause restructuring and breakup of
aggregates \cite{SoosJCIS08}, a phenomena macroscopically expressed in
shear thinning in dense suspensions \cite{ZacconePRL2011}. Breakup of
small aggregates due to hydrodynamic stress in turbulence is of high
relevance to various applications, e.g. processing of industrial
colloids, nanomaterials, wastewaters, and sedimentation of marine snow
\cite{SoosJCIS08,burdjackson09}. In a mean field situation,
aggregation-breakup dynamics are described by the Smoluchowski
equation. Defining $n_\xi(t)=N_\xi(t)/N_0$ where $N_\xi(t)$ is the
number concentration of aggregates consisting of $\xi$ primary particles,
and $N_0=\int_0^\infty d\xi\ \xi N_\xi(t)$,
the Smoluchowski equation reads as
\begin{align}
&\dot n_\xi(t)=-f_\xi n_\xi(t)+\int_\xi^\infty d\xi'\ g_{\xi,\xi'}f_{\xi'}n_{\xi'}(t) 
	+\frac{3\phi}{4\pi}  \left[\frac{1}{2}{\int_0^\xi} \right. \nonumber\\
&	\left. 
	 d\xi' k_{\xi',\xi\!-\!\xi'}n_{\xi'}(t)n_{\xi\!-\!\xi'}(t)
	- n_\xi(t)\int_0^\infty d\xi' k_{\xi,\xi'}n_{\xi'}(t) \right],
\label{Smol01}
\end{align}
where $\phi=\frac{4}{3}\pi a_p^3N_0$ is the solid volume fraction,
$a_p$ is the radius of the primary particle (assumed monodisperse and
spherical), and $k_{\xi,\xi'}$ is the aggregation rate. Breakup is
  accounted for by the fragmentation rate $f_\xi$
  $g_{\xi,\xi'}$. Determining these functions is not easy and despite
  considerable efforts \citep{babler_jfm08} a basic understanding of
  breakup dynamics is still lacking. A reason for this is the complex
  role of turbulence and the way it generates fluctuating stress to
  which an aggregate is exposed to.\\ The main issue we investigate in
  this paper is how to define and measure the fragmentation rate
  $f_\xi$ in a turbulent flow. The outcomes of our analysis are
  manifold: i) a Lagrangian and an equivalent Eulerian definition of
  the fragmentation rate can be derived, that fall off to zero in the
  limit of small aggregate mass, while they have a power-law behavior
  for large masses; ii) the power-law tail description is crucial to
  obtain a steady-state aggregate mass distribution when considering
  the full aggregation/breakup dynamics; iii) turbulent fluctuations
  allow for a broad asymptotic mass distribution, while a much
  narrower distribution of aggregates is obtained in the laminar
  case.\\ We adopt the simplest possible framework
  \cite{babler_jfm08}, and consider a dilute suspension of aggregates
  in a stationary homogeneous and isotropic turbulent flow.  We
  consider very small aggregates - much smaller then the Kolmogorov
  scale of the flow, in the range of $25$ to $100$ $\mu m$ for typical
  turbulent flows, with negligible inertia. Further, we assume that
  aggregates concentration is such that they do not modify the
  flow. Hence, their evolution is identical to that of passive
  point-like particles (unless extreme deviations from a spherical
  shape are present, e.g. elongated fibers). Moreover, the aggregates
  are brittle and breakup occurs instantaneously once being subject to
  a hydrodynamic stress that exceeds a critical value
  \cite{SonntagRussel1986shear}. For small and inertialess aggregates,
  the hydrodynamic stress exerted by the flow is
  $\sim\!\!\mu(\varepsilon/\nu)^{1/2}$, where $\mu$ and $\nu$ are the
  dynamic and kinematic viscosity, respectively and $\varepsilon$ is
  the local energy dissipation per unit mass. Thus, the key role is
  played by the turbulent velocity gradients across the aggregate
  which are known to possess strongly non-Gaussian, intermittent
  statistics \citep{PRL1991_Multifractal}.\\ Let $\epsCr(\xi)$ be the
  critical energy dissipation needed to break an aggregate of {\it
    mass} $\xi$. In the simplest case of a laminar flow, $\epsCr$
  relates to the critical shear rate for breakup as
  $G_\mathrm{cr}\sim(\epsCr/\nu)^{1/2}$. Earlier works
  \cite{SonntagRussel1986shear,Max2010} support the existence of a
  constituent power-law relation for $\epsCr$ implying that larger
  aggregates break at a lower stress than smaller ones: $\epsCr(\xi) =
  \langle\eps\rangle (\xi/\xi_s)^{-1/q}$ where the exponent $q$ is
  related to the aggregate structure and $\xi_s$ is the characteristic
  aggregate mass. An equivalent relation exists for small droplets
  breaking at critical capillary number, $\epsCr\sim
  \sigma^2/(\mu\rho\xi^{2/3})$ (here $\xi$ is the droplet volume, and
  $\sigma$ the interfacial energy). Hence, $f_\xi$ can be equally
  formulated in terms of a critical dissipation $f_{\epsCr(\xi)}$.
  \\ In this contribution we propose to define the
  fragmentation rate $f_\xi$ or $f_{\epsCr}$ in terms of a {\it first
    exit-time} statistics. This amounts to measure the fragmentation
  rate using the distribution of the time necessary to observe the
  first occurrence of a local hydrodynamic stress strong enough to
  break the aggregate. An operational formula of $f_{\epsCr}$ reads as
  follows: (i) seed homogeneously a turbulent, stationary flow with a
  given number of aggregates of mass $\xi$; (ii) neglect those
  aggregates in regions where the hydrodynamic stress is too high
  ($\eps > \epsCr$); (iii) from an initial time $t_0$, selected at
  random, follow the trajectory of each remaining aggregate until it
  breaks, and count the total number of breaking events in a given
  time interval $[\tau,\tau+d\tau]$. The time $\tau$ is the first {\it
    exit-time}: for an aggregate initally in a region with
  $\varepsilon < \epsCr$, $\tau$ is the time it takes to the
  hydrodynamic stress seen by the aggregate along its motion to cross
  the critical value $\epsCr$ at first opportunity, Fig. \ref{fig:1}.
The fragmentation rate is the inverse of the {\it mean} exit-time:
\begin{equation}\label{eq:exit}
f_{\epsCr} = \left[
\int_0^{\infty} \!d \tau\ \tau {\cal P}_{\epsCr}(\tau)\right]^{-1} = \frac{1}{\langle \tau(\epsCr) \rangle_{ex}}\,,
\end{equation}
where ${\cal P}_{\epsCr}(\tau)$ is the distribution of first exit-time
for a threshold $\epsCr$, and $\langle \cdot \rangle_{ex}$ is an
average over ${\cal P}_{\epsCr}(\tau)$. The definition in
Eq.~(\ref{eq:exit}) is certainly correct but difficult to implement
experimentally as it is needed to follow aggregate trajectories and
record the local energy dissipation, something still at the frontier
of nowadays experimental facilities \citep{Tsinober2001}. The question
thus arising is if we can obtain a proxy for the fragmentation rate
which is easier to measure.  Given a threshold for the hydrodynamic
stress $\epsCr$, one can measure the series $T_1,\ T_2,...$ of {\it
  diving-times}, namely the time lags between two consecutive events
of istantaneous stress crossing the threshold $\epsCr$ along the
aggregate motion, Fig.~\ref{fig:1}. In \citep{Loginov1985}, it was
proposed to estimate the fragmentation rate as the inverse of the mean
{\it diving-time}, $f_{\epsCr}^E = 1/\langle T(\epsCr) \rangle$.  An
important result is that $\langle T(\epsCr) \rangle$ can be obtained
using the Rice theorem for the mean number of crossing events per unit
time of a differentiable stochastic process across a threshold
\cite{Lundgren2006}. Hence,
\begin{align}
\label{eq:loginov}
f_{\epsCr}^E=\frac{1}{\langle T(\epsCr) \rangle} = \frac{\int_0^\infty d\dot\eps \ \dot\eps p_2(\epsCr,\dot\eps)}{\int_0^{\epsCr} d\eps\ p(\eps)}\,.
\end{align}
Here the numerator is the Rice formula giving the mean number of
crossings of $\epsCr$ in terms of the joint probability of dissipation
and its time derivative $p_2(\eps,\dot\eps)$; the denominator is the
measure of the total time spent in the region with $\eps
<\epsCr$. Notice that the integration in the numerator goes only on
positive values in order to consider only up-crossing of the threshold
$\epsCr$ \cite{Lundgren2006}. An obvious advantage of
Eq.~(\ref{eq:loginov}) is that it is {\it quasi-Eulerian}: it does not
require to follow trajectories, since it depends only the spatial
distribution of dissipation and of its first time derivative in the
flow. Expressions (\ref{eq:exit}) and (\ref{eq:loginov}) are not
strictly equivalent. A direct calculation of the mean exit-time in
terms of the distribution of diving-times gives indeed $ \langle
\tau(\epsCr) \rangle_{ex} = \langle T^2(\epsCr)\rangle/[ 2 \langle
  T(\epsCr)\rangle]$, which relates the mean exit-time to moments of
the diving-time.\\
Once a definition from first principles is set-up,
we proceed to measure the fragmentation rate for aggregates convected as
passive point particles in a statistically homogeneous and isotropic
turbulent flow, at Reynolds number $Re_\lambda \simeq 400$. Details on
the Direct Numerical Simulations (DNS) of Navier-Stokes equations with
$2048^3$ grid points and Lagrangian particles are in
\cite{BiferaleJFM2010a}. The present analysis is obtained averaging
over $6\times 10^5$ trajectories, recorded every $0.05 \tau_{\eta}$,
where $\tau_{\eta}$ is the Kolmogorov time of the
flow.\\ Figure~\ref{fig:2} shows the fragmentation rate measured from
the DNS data following the evolution of the velocity gradients along
aggregate trajectories. The exit-time measurement (\ref{eq:exit}) and
its Eulerian proxy (\ref{eq:loginov}) show a remarkable, non trivial
behavior for small values of the critical threshold, i.e. for large
aggregate mass. In this region, there is a competing effect between
the easiness in breaking a large aggregate and the difficulty to
observe a large aggregate existing in a region of low energy
dissipation. As a result, the estimated fragmentation rate develops a
quasi power-law behavior for small thresholds. On the other hand, for
large thresholds the super exponential fall off is expected. It is the
realm of very small aggregates that are broken only by large energy
dissipation bursts. The exit-time (\ref{eq:exit}) and diving time
(\ref{eq:loginov}) measurements are very close and we therefore
consider the latter a very good proxy of the former, the real
fragmentation rate. The main advantage of the estimate
(\ref{eq:loginov}) is the very high statistical confidence that can be
obtained since it is a quasi-Eulerian quantity. Moreover it gives a
reliable estimate also in the region of large thresholds where the
convergence of exit-time statistics, requiring very long aggregate
trajectories, is difficult to obtain.\\
Starting from Eq.~(\ref{eq:loginov}), simple models can be proposed
for the statistics of $\dot \eps$, whose direct measure requires a
very high sampling frequency along Lagrangian paths. First, as it
appears from the bottom inset of Fig.~\ref{fig:2}, the dissipation and
its time derivate are significantly correlated. Scaling on dimensional
grounds suggests $\dot\eps\sim \eps/\tau_{\eta}(\eps)$ where
$\tau_{\eta}(\eps)\sim(\nu/\eps)^{1/2}$ is the local Kolmogorov
time. It follows that the joint PDF $p_2(\eps,\dot \eps)$ can be
estimated as $p_2(\eps,\dot\eps)=\frac{1}{2}p(\eps)
\delta(|\dot\eps|-\eps/\tau_{\eta}(\eps)) $ where $p(\eps)$ is the
probability density of energy dissipation. Prefactor $1/2$ appears
since for a stationary process $\dot\eps$ is positive or negative with
equal probability. Plugging this expression in Eq.~(\ref{eq:loginov})
gives
\begin{align}\label{closure}
f_{\epsCr}^I=\frac{\frac{1}{2} \epsCr p(\epsCr)/\tau_{\eta}(\epsCr)}{\int_0^{\epsCr} d\eps\ p(\eps)}\,.
\end{align}
We refer to it as \textit{Closure I}. A different approach was
proposed in \citep{babler_jfm08}. It assumes that active regions in
the flow where $\eps>\epsCr$ engulf the aggregates at
a rate $\sim 1/\tau_\eta(\eps)$, which results in
\begin{align}\label{babler}
f_{\epsCr}^{II}=\frac{{\int_{\epsCr}^\infty
    d\eps\ p(\eps)/\tau_{\eta}(\eps) }}{\int_0^{\epsCr}d\eps\ p(\eps)}\,.
\end{align}
We refer to it as \textit{Closure II}. Both models share the advantage
of being fully Eulerian and based on the spatial distribution of the
energy dissipation only. In Figure~\ref{fig:2}, the fragmentation
rates obtained from the closures I and II are also shown. Both of them
reproduce the correct behavior for large values of the critical
dissipation but deviate for small ones.
\begin{figure}
\includegraphics[width=75mm]{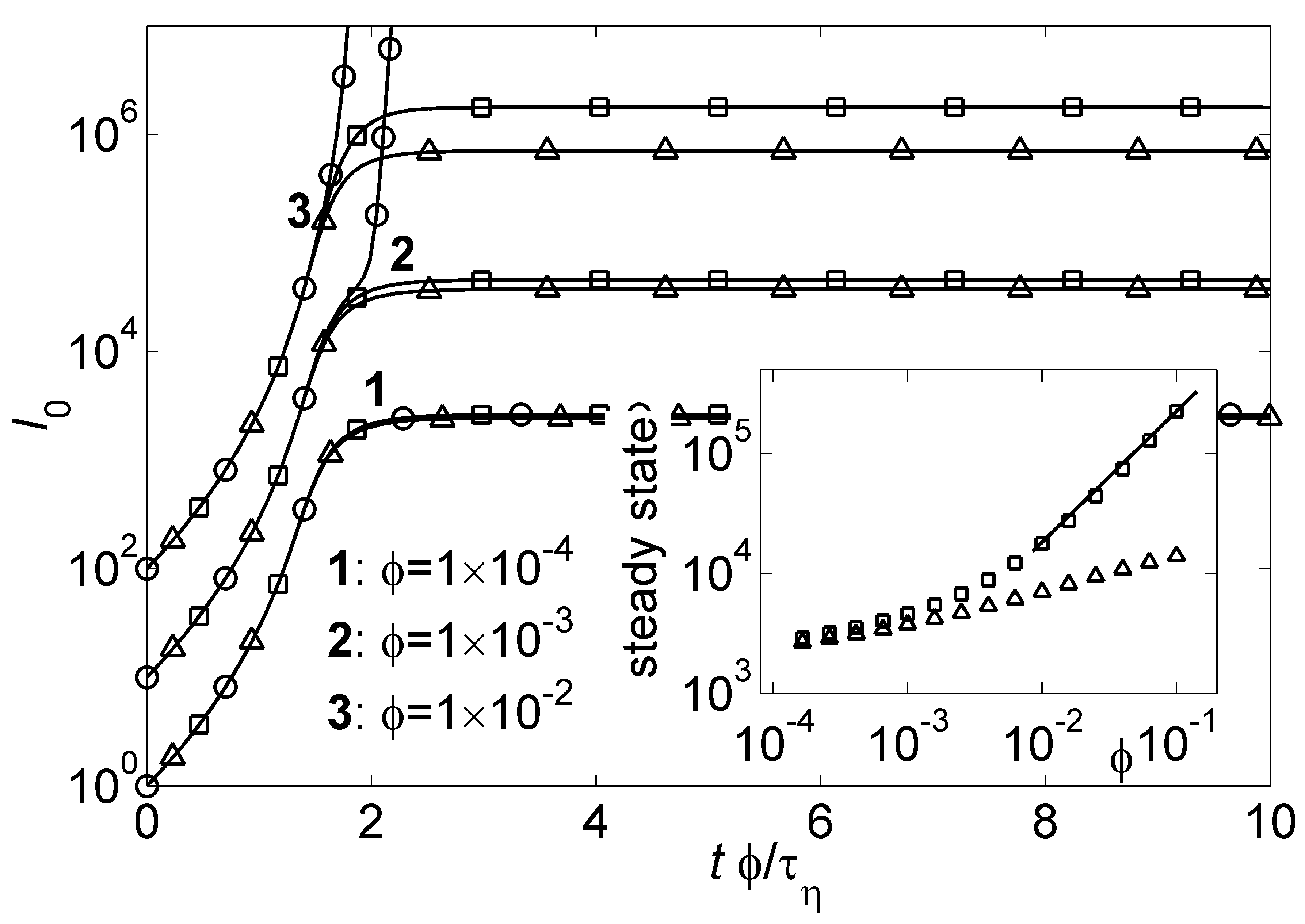}
\caption{\label{fig:3} Time evolution of $I_0$ for $d_f=2.4$ and
  $q=0.36$ with breakup rates $f_{\epsCr}^E$ (squares);
  $f_{\epsCr}^{I}$ (circles); $f_{\epsCr}^{II}$ (triangles). Runs with
  different solid volume fractions $\phi$ are shifted. Inset: $I_0$ at
  steady state as a function of $\phi$ (same symbols). At large
  $\phi$, the model with fragmentation (\ref{eq:loginov}) relaxes to
  the predicted scaling curve $I_0 \sim\phi^{1/(1+\chi/q-3/d_f)}$,
  given by the solid line. Here, $\xi_s=10^4$ implying that
  $a/a_p\sim\xi_s^{1/d_f}\approx 50$ is the value of the characteristic
  aggregate size.}
\end{figure}
 The reason for such discrepancy is made clear in the top inset of
 Fig.~\ref{fig:2}, where the numerator of Eqs.~(\ref{closure}) and
 (\ref{babler}) are shown. It appears that, for small values of the
 critical stress $\epsCr$, $f_{\epsCr}^{I}$ underestimates the number
 of breakup events, while the numerator of $f_{\epsCr}^{II}$ saturates
 to a constant value.\\ Since in experiments breakup {\it a fortiori}
 takes place together with aggregate recombination, we explore how the
 actual fragmentation rate Eq.~(\ref{eq:loginov}) and the two
 closures, Eqs.~(\ref{closure})-(\ref{babler}), influence the time
 evolution of an ensemble of aggregates $n_\xi(t)$. At this purpose,
 the Smoluchowski equation (\ref{Smol01}), subject to the initial
 condition $n_\xi(0)=\delta(\xi-1)$, is evolved in time. To model
 aggregation we use the classical Saffman-Turner expression
 $k_{\xi,\xi'}=D_0/\tau_\eta (\xi^{1/d_f}+\xi'^{1/d_f})^3$, where
 $D_0$ is an $~O(1)$ constant and $d_f$ is the fractal dimension that
 relates the collision radius of an aggregate to its mass,
 $a/a_p=\xi^{1/d_f}$. Breakup is assumed to be binary and symmetric,
 $g_{\xi,\xi'}=2\delta(\xi-\xi'/2)$, which despite its simpleness
 represents well the quality of the evolution. To quantify our
 findings we consider the second moment
 observable readily accessible from static light scattering
 \citep{SonntagRussel1986shear}. Fig.~\ref{fig:3} shows the time
 evolution of $I_0$ for typical values of $d_f$ and $q$ found in
 turbulent aggregation of colloids \citep{SoosJCIS08}.  After an
 initial growth period, curves obtained with Eqs.~(\ref{eq:loginov})
 and (\ref{babler}) both relax to a steady state. At small solid
 volume fraction $\phi$, the two models nearly overlap, whereas at
 larger $\phi$ Closure II underestimates $I_0$. This behavior is
 confirmed in the inset of Fig.~\ref{fig:3} that shows $I_0$ at steady
 state for both models and for different values of $\phi$. Clearly, at
 large $\phi$ other phenomena, i.~e. modulation of the flow due to the
 particles, may occur which, however, is beyond the scope of the
 present work.  Closure I shows a very different behavior. At small
 $\phi$, an evolution similar to previous cases is observed. However,
 by increasing $\phi$ a drastic change appears and $I_0$ diverges, a
 direct consequence of the presence of a maximum in the shape of
 $f_{\epsCr}^{I}$, see Fig.~\ref{fig:2}.\\ The time evolution of the
 number concentration $n_\xi(t)$, governed by Eq.~(\ref{eq:loginov}),
 is further examined in Fig.~\ref{fig:4}. Here, we compare the steady
 distribution obtained in the turbulent flow with that of a laminar
 flow for different values of the solid volume fraction. In the
 turbulent case, $n_\xi(t)$ rapidly grows and reaches a stationary
 state whose peak mode is controlled by the magnitude of aggregation,
 i.e. the solid volume fraction $\phi$. Increasing $\phi$ causes the
 mode to broaden and to shift to the right as aggregation gets more
 pronounced. The distribution thus gradually moves into the region
 where $f_{\epsCr}$ assumes power-law behavior. A numerical fit of the
 left fragmentation tail in Fig.~\ref{fig:2} gives $f_\xi
 \sim\xi^{\chi/q}$, with $\chi=0.42 \pm 0.02$ (dashed curve in
 Fig.~\ref{fig:4}). Using this latter expression in
 Eq.~(\ref{Smol01}), one can derive a scaling relation for integral
 quantities of $n_\xi(t)$ at steady state \cite{Thes_jcis}. Such
 scaling is reported in the inset of Fig.~\ref{fig:3}.  On the other
 hand, in the laminar case where a uniform shear rate governs the
 breakup, the steady state distribution is much narrower and shows
 multiple resonant modes. These are due to the sharp onset of breakup
 once the aggregates grow larger than the characteristic aggregate
 mass $\xi_s$. These results clearly demonstrate the strong influence
 of turbulent fluctuations on the statistically stationary mass
 distribution function.\\
We have presented a study of the fragmentation rate of small and
diluted aggregates in turbulent flows at high Reynolds number. We have
introduced a novel expression for the fragmentation rate in terms of
the {\it exit-time} statistics, which is a natural way of measuring
first-order rate process. Also, a purely Eulerian proxy based on
Eq.~(\ref{eq:loginov}) provides a very good approximation to the
actual fragmentation rate measured from our DNS. Remarkably, a steady
state in the full breakup-aggregation process is crucially determined
by the left tail of the fragmentation rate, i.e. by events of low
energy dissipation. Our investigation puts the basis for many
developments, such as the stability of the Smoluchowski evolution
using the measured fragmentation rates in experiments, and the
extension to the case of {\it inertial} aggregates. In such case, the
correlation between the hydrodynamic shear and the Stokes drag may
result in a non-trivial breakup rate dependency on the degree of
inertia. Future work aims to introduce spatial fluctuations in the
mass distribution caused by local breakup, a research path still
poorly explored.

EU-COST action MP0806 is kindly acknowledged. L.B. and A.L. thank the
DEISA Extreme Computing Initiative and CINECA (Italy) for technical
support.


\begin{thebibliography}{99}

\bibitem{BiferaleJFM2010b}
G.~Falkovich, A.~Fouxon, and M.~G. Stepanov,
Nature (London) \textbf{419}, 151 (2002).
J.~Bec, {\it et al.} J. Fluid Mech. \textbf{646}, 527 (2010).
J.~Chun, {\it et al.} J. Fluid Mech, \textbf{536}, 219 (2005).

\bibitem{BiferaleJFM2010a}
J.~Bec, {\it et al.} J. Fluid Mech. \textbf{645}, 497 (2010).

\bibitem{SoosJCIS08}
M.~Soos {\it et al.} J. Colloid Interface Sci. \textbf{319} 577 (2008).
A.~Zaccone {\it et al.} Phys. Rev. E \textbf{79}, 061401 (2009).
V.~Becker, {\it et al.} J. Colloid Interface Sci. \textbf{339}, 362 (2009).

\bibitem{ZacconePRL2011}
X.~Cheng,{\it et al.} Science \textbf{333}, 1276 (2011)
E.~Brown,{\it et al.} Nature Mater. \textbf{9}, 220 (2010)

\bibitem{burdjackson09}
A.~B. Burd and G.~A. Jackson,
Annu. Rev. Mar. Sci. \textbf{1}, 65 (2009).
R.~Wengeler, {\it et al.} Langmuir \textbf{23}, 4148 (2007).
C.~Selomulya,  {\it et al.} Langmuir \textbf{18}, 1974 (2002).

\bibitem{babler_jfm08} M.~U. Babler, {\it et al.}, J. Fluid
  Mech. \textbf{612}, 261 (2008).

\bibitem{SonntagRussel1986shear}
R.~C. Sonntag and W.~B. Russel,
J. Colloid Interface Sci. \textbf{113}, 399 (1986).

\bibitem{PRL1991_Multifractal}
R.~Benzi,{\it et al.},
Phys. Rev. Lett. \textbf{67}, 2299 (1991).

\bibitem{Max2010}
M.~L. Eggersdorfer,{\it et al.}
J. Colloid Interface Sci. \textbf{342}, 261 (2010).
Y.~M. Harshe, M.~Lattuada, and M.~Soos,
Langmuir \textbf{27}, 5739 (2011).

\bibitem{Tsinober2001}
B.~Luthi, A.~Tsinober, and W.~Kinzelbach,
J. Fluid Mech. {\bf 528}, 87 (2005).

\bibitem{Loginov1985}
V.~I. Loginov,
J. Applied Mech. Tech. Phys. \textbf{26}, 509 (1985).

\bibitem{Lundgren2006} 
G.~Lindgren, \emph{Lectures on stationary stochastic processes} (Lund University, 2006)

\bibitem{Meakin}
F. Family, P. Meakin, and J.~M. Deutch, Phys. Rev. Lett. \textbf{57}, 727 (1986).
C.~M. Sorensen, H.~X. Zhang, and T.~W. Taylor, Phys. Rev. Lett. \textbf{59}, 363 (1987).

\bibitem{Thes_jcis}
M.~U. Babler, M.~Morbidelli, J. Colloid Interface Sci. \textbf{316}, 428 (2007).

\end{thebibliography}
\end{document}